# The influence of pressure waves and flame collisions on the development and evolution of tulip flames

Chengeng Qian [a], Mikhail A. Liberman [b*]

[a] *Aviation Key Laboratory of Science and Technology on High Speed and High Reynolds Number, Shenyang Key Laboratory of Computational Fluid Dynamics, Aerodynamic Force Research AVIC Aerodynamics Research Institute, Shenyang 110034, China*

[b] *Nordita, KTH Royal Institute of Technology and Stockholm University, Hannes Alfvéns väg 12, 114 21 Stockholm, Sweden*

## Abstract

The effects of pressure waves-flame collisions and tube aspect ratio on flame evolution and the formation of tulip and distorted tulip flames were investigated using numerical simulations of the fully compressible Navier-Stokes equations coupled with a detailed chemical model for a stoichiometric hydrogen-air mixture. It is shown that: (1) the rarefaction wave generated by the decelerating flame in the unburned gas is the primary physical process leading to the flame front inversion and the tulip flame formation, (2) the flame front instabilities (Darrieus-Landau or Rayleigh-Taylor) do not participate in the formation of the tulip flame, since the time of the flame front inversion due to the rarefaction wave is considerably shorter than the characteristic times of the development of instabilities with wavelengths of the order of the tube width. The first rarefaction wave in the unburned gas mixture is generated after the flame skirt touches the tube walls and the flame is slowed down due to the reduction in flame surface area. The collision of the flame with the pressure waves reflected from the closed end of the tube leads to a faster and more pronounced formation of a tulip-shaped flame. In later stages, flame collisions with pressure waves can lead to the formation of distorted tulip flames due to short-wavelength Rayleigh-Taylor instability of the flame front. Because flame acceleration and deceleration occur much faster in 3D flames than in 2D flames, tulip flame formation also occurs much faster in 3D flames than in 2D flames.

**Novelty and Significance**

The novelty of this study lies in the convincing evidence that rarefaction waves generated during the flame deceleration phases play a key role in flame front inversion and tulip flame formation. A rarefaction wave can be amplified by flame collisions with pressure waves reflected from the closed end of the tube. Tulip flame formation is shown not to involve intrinsic flame instabilities (Darrieus-Landau and Rayleigh-Taylor). The time of flame front inversion under the action of the rarefaction wave is much shorter than the characteristic times of flame instability. Tulip flame collisions with pressure waves can lead to short-wavelength Rayleigh-Taylor instability of the flame front, resulting in the formation of a distorted tulip flame.

*Significance*: Flame propagation in tubes is the basic physical-chemical platform for the analysis of more complex engineering and technological problems, and for the development of analytical methods and numerical models important to many scientific and engineering combustion processes. The early stages of flame propagation are of particular interest as they influence the flame acceleration/deceleration that largely determine subsequent combustion regimes.

# 1. Introduction

Flame propagation in tubes is the basic fundamental physical-chemical platform for the analysis of more complex engineering and technological problems, and for the development of analytical methods and numerical models important to many scientific and engineering combustion problem. These include explosion safety in confined and unconfined spaces, combustion efficiency, combustion processes in gas turbines and internal combustion engines, pollutant emissions, etc. The early stages of flame propagation are of particular interest as they influence the flame acceleration/deceleration that largely determine subsequent combustion regimes.

It is well known that a flame ignited near the closed end of a tube and propagating to the opposite closed or open end suddenly slows down, and the shape of the flame front rapidly changes from a convex shape with the tip pointing forward to a concave shape with the tip pointing backward. This phenomenon was first photographed in experimental studies by Ellis [1, 2] almost a century ago and was subsequently named the "tulip flame" phenomenon [3]. Many subsequent experimental studies have shown that tulip flame formation is remarkably robust to all combustible mixtures and downstream conditions: open or closed tube ends, tube shapes, etc. There is a large body of literature on experimental, theoretical and numerical studies conducted with the purpose of explaining the physical origin (mechanism) of tulip flame formation. Different scenarios of tulip flame formation have been considered, including flame collision with a shock wave [4, 5], Darrieus-Landau (DL) [6, 16] and Rayleigh-Taylor (RT) [17] flame front instabilities. Flame-vortex interaction has been considered by many authors [7, 18--26] as a seemingly plausible scenario for flame front inversion, although no evidence has been provided that this is actually the case. The reader can find a long history of tulip flame research and relevant references in the reviews by Dunn-Rankin [27] and Searby [28] and in more recent publications [29, 30, 31]. However, despite numerous experimental, theoretical and numerical studies, the mechanism of the tulip flame formation, including the role of pressure waves, remains one of the major unsolved fundamental problems in combustion science. This is partly because the meaning of tulip flame formation is not entirely clear. Indeed, the inversion of the flame front can be caused by the flame collision with a shock wave or by the flame front instabilities. To avoid misunderstandings, we refer to the tulip flame phenomenon as the rapid inversion of the flame front during the deceleration of the finger flame, in which the inherent instabilities of the flame front are not involved, which is consistent with the conclusion reached in the experimental studies of Ponizy et al [32].

Recent work by Liberman et al [29] has shown that the formation of tulip flames is a purely gas-dynamic phenomenon in the sense that it is faster than the development of flame front instabilities. It has been shown that the rarefaction wave generated by the flame during the deceleration phase, when the flame skirt touches the sidewalls, resulting in a reduction of the flame surface area, which in turn reduces the flame speed, is the principle physical mechanism responsible for flame front inversion and tulip flame formation. The mechanism of flame front inversion and tulip flame formation is remarkably simple and works as follows [29]. In the theoretical model of a thin flame [30, 33], the flame front is considered as a discontinuous surface between the unburned and burned gases. Each point of the flame front propagates at the velocity equal to the sum of the laminar flame velocity at which the flame propagates relative to the unburned fuel and the velocity of the unburned gas ahead of that point at which the unburned gas entrains that portion of the flame front. After ignition, the flame front surface increases from a small spark to an expanding hemisphere and then to a finger-shaped flame [17]. During these stages, the surface area of the flame increases and the flame velocity increases as the flame surface area increases. The accelerating flame acts like a semi-transparent piston, creating pressure waves and a flow of unburned gas ahead of the flame. The axial velocity profile in unburned gas is nearly convex: the velocity is maximum near the tube axis, decreases in the boundary layer, and disappears at the tube sidewalls. When the flame skirt touches the tube sidewalls, the lateral portions of the flame skirt are extinguished, the flame surface area decreases, resulting in a decrease in flame velocity. At this stage, the flame continues to decelerate as the sides of the flame skirt collapse against the tube walls, and rarefaction waves are continuously generated by the decelerating flame. In the reference frame of the unburned gas, the decelerating flame acts as a "piston" that begins to move out of the tube with acceleration. It is well known [34] that such a piston creates a simple rarefaction wave. The front of the rarefaction wave propagates forward through the unburned gas at the speed of sound, creating a reverse flow in the unburned gas. The gas velocity is directed everywhere to the left (reverse flow) and monotonically decreases in magnitude in the positive direction toward the front of the rarefaction wave. The superposition of the unburned gas flow created by the flame during the acceleration phases and the reverse flow created by the rarefaction wave leads to a decrease in the unburned gas velocity in the close vicinity of the flame front and, consequently, to an increase of the boundary layer thickness. Unlike the classical rarefaction wave produced by a flat piston, the decelerating piston (flame front) has a convex shape. It is obvious that in this case the velocity of the reverse flow created by the rarefaction wave is maximum at the tube axis and decreases towards the tube walls. The resulting axial velocity profile of the unburned gas in the immediate vicinity of the flame front changes from a convex to a concave shape in the bulk flow between the boundary layers, where the velocity decreases and disappears at the tube sidewalls. Since the unburned gas axial velocity profile is tulip-shaped in the vicinity of the flame front, the flame front also becomes tulip-shaped. According to the thin flame model, the flame front takes on a tulip shape, with the half-thickness of the tulip tongues equal to the thickness of the boundary layer in the flow immediately ahead of the flame. The results of the thin flame model appeared to be in a good agreement with 2D and 3D simulations [29], which used a detailed chemical-kinetic model for stoichiometric hydrogen-air mixtures, thus accounting for the true thickness of the flame front.

It should be emphasized that the mechanism of the tulip flame formation due to the rarefaction wave is much faster than the characteristic time of the intrinsic instabilities of the flame front, which means that these instabilities are not involved in the process in agreement with experimental studies [32]. For example, the time of the Darrieus-Landau instability development can be estimated as $\tau_{DL} = 1/\sigma_{DL}$, where

$$\sigma_{DL} = kU_{fL} \frac{\Theta}{\Theta+1}\left[\sqrt{\Theta+1-1/\Theta} -1\right] \approx kU_{fL}\sqrt{\Theta} \qquad (1)$$

is the increment of the DL instability, $U_{fL}$ is the laminar flame speed, $\Theta = \rho_u/\rho_b$ is expansion coefficient, the ratio of unburned to burned gas densities, $k$ is the wave number. The time of establishing the reverse flow by a rarefaction wave can be estimated as $\tau_{RW} \approx D/a_s$, where $a_s$ is a speed of sound. For the wavelength of perturbation $\lambda \sim D$, where $D$ is the tube width, we obtain

$$\tau_{RW}/\tau_{DL} \approx \pi\sqrt{\Theta}\left(\frac{U_{fL}}{a_s}\right) \ll 1. \qquad (2)$$

The laminar flame velocity $U_{fL}$ is much less than the speed of sound $U_{fL}/a_s \sim 10^{-3}$, and the condition (2) is valid for all combustible gas mixtures. Taking for hydrogen-air flame $U_{fL} \approx 2.36\,m/s$, sound speed $a_s \approx 408\,m/s$, $\Theta = 7.8$, we obtain $\tau_{RW}/\tau_{DL} \sim 0.1$. It is interesting to note that the fact that the formation of tulip flames occurs much faster than the characteristic time of the DL instability has been established long ago in the numerical studies by Gonzalez et al [9] and Dunn-Rankin et al [8], but the mechanism of tulip flame formation remained unclear.

In this paper, we examine the role of pressure waves in the formation of the tulip flame and in subsequent stages when a distorted tulip flame is formed. We show that the collision of the flame with pressure waves reflected from the opposite closed end of the tube can amplify the intensity of the rarefaction wave, leading to a faster inversion of the flame front, and in later stages can cause the short-wavelength ($\lambda \ll D$) of the RT instability when the flame collides with pressure waves, resulting in the formation of a distorted tulip flame [24, 26].

## 2. Numerical models and physical parameters

### 2.1 Two-dimensional DNS

The two-dimensional computational domains, that were modeled, are the rectangular channels of width $D = 1\,cm$, the aspect ratio of $L/D = 6, 12, 18$ with both ends closed and a rectangular channel of width $D = 1\,cm$ with an open right end. The two-dimensional, time-dependent, reactive compressible Navier-Stokes equations including molecular diffusion, thermal conduction and viscosity are solved using a fifth-order WENO finite-difference method [36]. The effect of heat loss can be neglected due to the short reaction time and high flame velocity, so no-slip adiabatic boundary conditions are used on the tube walls. The flame is ignited by a small semi-circular pocket of hot, burned gas at $T_b = 2350\,K$ and pressure $P_0 = 1$ atm with the radius 3mm, at center $x = y = 0$ at the left closed end of the tube; the $x$ coordinate is along the tube, $y = 0$ is the tube axis, the tube sidewalls are $y = \pm D/2$. A detailed chemical kinetic model for a stoichiometric hydrogen–air mixture consisting of 19 reactions and 9 species developed by Kéromnès et al. [37] was implemented in simulations. This chemical model has been extensively tested against experimental data and found to accurately predict ignition delay and laminar flame velocity over a wide range of pressures (1–70 bar), temperatures (900–2500 K), and equivalence ratio. (0.1-4.0). The initial conditions are $P_0 = 1\,atm$, $T_0 = 298\,K$. During the formation of a tulip flame, the pressure increases from the initial value of $P_0 = 1\,atm$ up to $\approx 2\,atm$ for channels with high aspect ratio ($L/D = 12, 18$) and up to $\sim 4.0\,atm$ in the short channel $L/D = 6$.

The resolution and convergence tests (grid independence) similar to those in our previous publications [38, 39] were thoroughly performed to ensure that the resolution is adequate to capture details of the problem in question and to avoid computational artifacts. The governing equations are

$$\frac{\partial \rho}{\partial t} + \frac{\partial (\rho u_i)}{\partial x_i} = 0, \tag{3}$$

$$\frac{\partial (\rho u_i)}{\partial t} + \frac{\partial (P\delta_{ij} + \rho u_i^2)}{\partial x_j} = \frac{\partial \tau_{ij}}{\partial x_j}, \tag{4}$$

$$\frac{\partial (\rho E)}{\partial t} + \frac{\partial \left[(\rho E + P)u_i\right]}{\partial x_i} = \frac{\partial (\tau_{ij} u_i)}{\partial x_i} - \frac{\partial q_i}{\partial x_i}, \tag{5}$$

$$\frac{\partial \rho Y_k}{\partial t} + \frac{\partial \rho u_i Y_k}{\partial x_i} = \frac{\partial}{\partial x_i}\left(\rho V_{ik} Y_k\right) + \dot{\omega}_k, \tag{6}$$

$$P = \rho R_B T \left( \sum_{i=1}^{N_s} \frac{Y_i}{W_i} \right), \tag{7}$$

Here $\rho$, $u_i$, $T$, $P$, $E$, $\tau_{ij}$, $q_i$ are density, velocity components, temperature, pressure, specific total energy, viscosity stress, heat flux mass, $R_B$ is the universal gas constant. $Y_i, W_i, V_{i,j}$ are mass fraction, molar mass and diffusion velocity of species $i$. The viscosity and thermal conductivity of pure-species and mixture are calculated by the Chapman-Enskog expression Byron Bird et al. [40] and a semi-empirical formula Wilke [41].

The reaction rate of species $i$ is determined as

$$\dot{\omega}_k = W_i \sum_{j=1}^{N_r} (v''_{jk} - v'_{jk}) \cdot \left( k_{f,j} \prod_{k=1}^{N_s} \left(\frac{\rho Y_k}{W_k}\right)^{v'_{jk}} - k_{b,j} \prod_{k=1}^{N_s} \left(\frac{\rho Y_k}{W_k}\right)^{v''_{jk}} \right) \tag{8}$$

where $v'_{jk}$ and $v''_{jk}$ are stoichiometric coefficients of species $k$ of the reactant and product sides of reaction $j$.

The input parameters describing a stoichiometric mixture of hydrogen/air are: initial pressure $P_0 = 1 atm$, initial temperature $T_0 = 298K$, initial density $\rho_0 = 8.5 \cdot 10^{-4} g/cm^3$, laminar flame velocity 2.36m/s, laminar flame thickness $L_f = 0.0325 cm$, adiabatic flame temperature $T_b = 2350K$ expansion coefficient $\Theta = \rho_u / \rho_b = 7.8$, specific heat ratio $\gamma = C_P / C_V = 1.399$, sound speed $a_s = 408.77 m/s$.

## 2.2 Three-dimensional LES

The three-dimensional case was modeling using Large Eddy Simulation (LES). The thickened flame model [42] was used, when the flame front is thickened artificially without changing the laminar flame velocity. A comparison of the flame dynamics obtained with DNS for a flame propagating in a two-dimensional channel with the flame dynamics obtained with LES has shown that the results of DNS and LES are reasonably close [29].

In LES, all quantities are filtered in the spectral space by applying filter to the governing equations. Since the flow is compressible, the mass-weighted Favre is introduced. The filtered governing equations are

$$\frac{\partial \bar{\rho}}{\partial t} + \frac{\partial \bar{\rho} \tilde{u}_i}{\partial x_i} = 0, \tag{9}$$

$$\frac{\partial \bar{\rho} \tilde{u}_i}{\partial t} + \frac{\partial}{\partial x_j}\left(\bar{\rho}\tilde{u}_i\tilde{u}_j + \bar{P}\delta_{ij}\right) = \frac{\partial \bar{\tau}_{ij}}{\partial x_j} - \frac{\partial \tau_{ij}^{sgs}}{\partial x_j}, \tag{10}$$

$$\frac{\partial \bar{\rho}\tilde{E}}{\partial t} + \frac{\partial}{\partial x_i}(\bar{\rho}\tilde{E} + \bar{P})\tilde{u}_i = \frac{\partial(\tilde{u}_i\bar{\tau}_{ij})}{\partial x_i} - \frac{\partial \bar{q}_i}{\partial x_i} - \frac{\partial}{\partial x_i} H_i^{sgs} + \frac{\partial}{\partial x_i}\sigma_{ij}^{sgs}, \tag{11}$$

$$\frac{\partial \bar{\rho}\tilde{Y}_k}{\partial t} + \frac{\partial}{\partial x_i}(\bar{\rho}\tilde{Y}_k\tilde{u}_i) = \frac{\partial}{\partial x_i}\left(\bar{\rho}\tilde{D}_i \frac{\partial \tilde{Y}_k}{\partial x_i}\right) - \frac{\partial \varphi_i^{sgs}}{\partial x_i} + \bar{\omega}_k \tag{12}$$

The unresolved fluxes $\sigma_{ij}^{sgs}$ are negligible, and the unresolved Reynolds stresses $\tau_{ij}^{sgs}$ are modeled according to the Boussinesq eddy viscosity assumption

$$\tau_{ij}^{sgs} - \frac{1}{3}\delta_{ij}\tau_{kk}^{sgs} = -\rho v_t \left(\frac{\partial \tilde{u}_i}{\partial x_j} + \frac{\partial \tilde{u}_j}{\partial x_i} - \frac{2}{3}\delta_{ij}\frac{\partial \tilde{u}_k}{\partial x_k}\right). \tag{13}$$

The sub-grid viscosity $v_t$ is calculated by wall-adapting local eddy viscosity model and the sub-grid heat flux is defined by Garnier et al. [43]

$$H_i^{sgs} = -\bar{\rho}\frac{v_t c_p}{\Pr_t}\frac{\partial \tilde{T}}{\partial x_i}, \qquad (14)$$

where $\Pr_t = 0.75$ sub-grid Prandtl number.

In simulations of turbulent reactive flow, the most important is the combustion model, which is used to approximate the filtered reaction rates. In the thickened flame model, the flame front is artificially thickened without changing the laminar flame velocity. The laminar flame velocity $U_{fL}$ and the flame thickness $L_f$ can be expressed as

$$U_{fL} \propto \sqrt{D_{th} A}, \quad L_f \propto \frac{D_{th}}{U_f} = \sqrt{D_{th}/A}, \qquad (15)$$

where $D_{th}$ and $A$ are thermal diffusivity and pre-exponential factor. If $D_{th}$ and $A$ are replaced by $FD_{th}$ and $A/F$, it is easy to check that $L_f$ increases by a factor $F$ while $U_{fL}$ remains unchanged. The wrinkling factor function $\varepsilon$ is introduced to simulate a flame front wrinkled by vortices at the subgrid scale. More details about simulations, resolution and convergence tests can be found in [29, 38]. For a fast hydrogen-air flame the heat loss to the walls is negligible, therefore the no-slip adiabatic reflecting boundary conditions at the tube walls were used [44]

$$\vec{u} = 0, \ \partial T/\partial \vec{n} = \partial Y_k/\partial \vec{n} = 0, \qquad (16)$$

where $\vec{n}$ is the normal to the wall (y-axis). To model a half-open tube, the non-reflecting outflow boundary condition is used to calculate the subsonic outflow [39].

In the 3D simulations we use a simplified one-step chemical model. For the irreversible global reaction

$$H_2 + 0.5 O_2 \Rightarrow H_2O. \qquad (17)$$

The reaction rate is taken in the form a one-step Arrhenius-type chemical kinetics. For pressures $P < 2\,\text{bar}$:

$$\dot{\omega} = dY_{H_2}/dt = A\exp\left(-\frac{E_a}{R_B T}\right)\left(\frac{\rho Y_{H_2}}{W_{H_2}}\right)\left(\frac{\rho Y_{O_2}}{W_{O_2}}\right), \tag{18}$$

and for pressures $P > 2\,\text{bar}$:

$$\dot{\omega} = dY_{H_2}/dt = A\exp\left(-\frac{E_a}{R_B T}\right)\left(\frac{\rho Y_{H_2}}{W_{H_2}}\right)^{0.9}\left(\frac{\rho Y_{O_2}}{W_{O_2}}\right)^{0.9}. \tag{19}$$

The reaction order is $n = 2$ for $P < 2\,\text{bar}$, and the reaction order is $n = 1.8$ for $P > 2\,\text{bar}$. It has been verified in [31] that the flame dynamics, temporal evolution of flame velocity and flame surface area obtained using the thickened flame model with the one-step chemical model are reasonably close to those obtained using the high-order DNS with the detailed chemical model.

The maximum pressure during the tulip flame formation in the case of a short tube with both ends closed is about 3 bar, resulting in the decrease in the thickness of the flame front from $L_f = 325\,\mu\text{m}$ to $L_f \simeq 106\,\mu\text{m}$. A uniform mesh with a resolution $\Delta x \simeq 12.5\,\mu\text{m}$, which corresponds to 28 grid points across the flame width at the beginning of the process and 9 grid points at maximum pressure for a tube closed at both ends, was used in the simulations. Thorough resolution and convergence (a grid independence) tests were performed in previous publications [29, 38] by varying the value of $\Delta x$ to ensure that the resolution is adequate to capture details of the problem in question and to avoid computational artefacts. The parameters used in simulations are shown in Table 2.

**Table 2.** Model parameters for simulating a stoichiometric hydrogen–air flame.

| Initial pressure | $P_0$ | 1.0 atm |
|---|---|---|
| Initial temperature | $T_0$ | 298 K |
| Initial density | $\rho_0$ | $8.5 \cdot 10^{-4}\,\text{g}/\text{cm}^3$ |
| Pre-exponential factor $P < 2\,\text{bar}$ | $A$ | $2.95 \cdot 10^{13}\,[cm^3/mol^3 \cdot s]$ |

| Pre-exponential factor $P > 2$ bar | $A$ | $2.1 \cdot 10^{12} \, [cm^{2.4} / mol^{0.4} \cdot s]$ |
|---|---|---|
| Activation energy | $E_a$ | $27 R_B T_0$ |
| Laminar flame velocity | $U_{fL}$ | 2.36 m/s |
| Laminar flame thickness | $L_f$ | 0.0325 cm |
| Adiabatic flame temperature | $T_b$ | 2500 K |
| Expansion coefficient ($\rho_u / \rho_b$) | $\Theta$ | 8.34 |
| Specific heat ratio | $\gamma = C_P / C_V$ | 1.399 |
| Sound speed | $a_s$ | 408.77 m/s |

## 3. Results

### *3.1 Two-dimensional tubes*

In this section we describe the dynamics of flames propagating in tubes with aspect ratios $\alpha = L/D = 6, 12, 18$ with both ends closed and in a semi-open tube (open right end). The initial stages of flame propagation after ignition near the left end: (1) spherical flame, (2) finger-shaped flame, have been described by many authors [8, 11, 17, 21, 26, 27]. Fig. 1(a) shows the calculated temporal evolution of the flame front velocities along the tube axis, $y = 0$, for the tubes with aspect ratios $\alpha = 6$ and $\alpha = 12$ with both ends closed. Fig. 1(b) shows the calculated temporal evolution of the flame front velocities along the tube axis for the tube with aspect ratio $\alpha = 18$ and in a semi-open tube. Also, Fig. 1 shows the pressure evolution immediately ahead of the flame front. It can be seen that the initial stages of flame acceleration up to the moment of 0.55ms are almost the same for the flame in closed tubes with different aspect ratios and for a semi-open tube. In Fig. 1(a), a slight decrease in flame velocity at 0.3-04ms can be seen for the tube with aspect ratio $\alpha = 6$, caused by the flame collision with the reflected pressure wave.

After the flame skirt touches the tube sidewalls at 0.55ms, the flame surface area decreases, resulting in a decrease in flame velocity. The subsequent dynamics of the flame will depend on the aspect ratio of the tube. The collision of the flame with the pressure wave reflected from the opposite closed end of the tube causes an additional decrease of the flame velocity, which can lead either to an intensification of the rarefaction wave effect or to the development of the Rayleigh-Taylor instability leading to distortion of the tulip-shaped flame. The time and number of flame collisions with pressure waves depends on the aspect ratio of the tube; the shorter the tube, the greater the number of collisions.

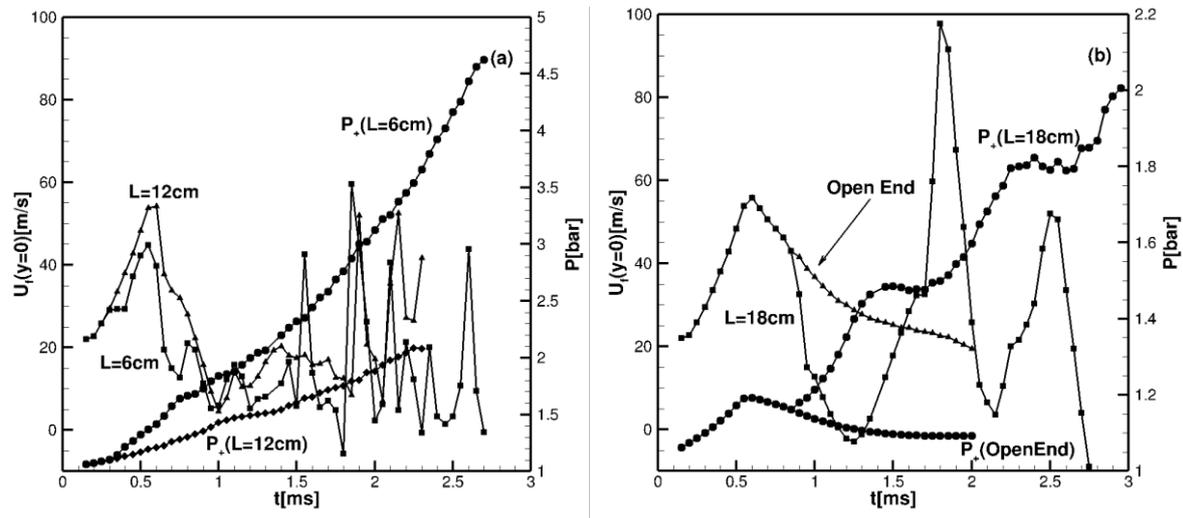

**Fig. 1: a)** The velocity of the flame front at the tube axis, $y=0$, computed for tubes $L/D=6, 12$; **b)** the same for the tube $L/D=18$ and the half-open tube. Also shown the pressure evolution just ahead of the flame front, $P_+$.

Figures 2(a, b) show the temporal evolution of the flame front velocities along the centerline, $y = 0$, and the pressure growth rate $dP_+ / dt$ at the centerline just ahead of the flame front, indicating the location of the pressure waves just before their collisions with the flame.

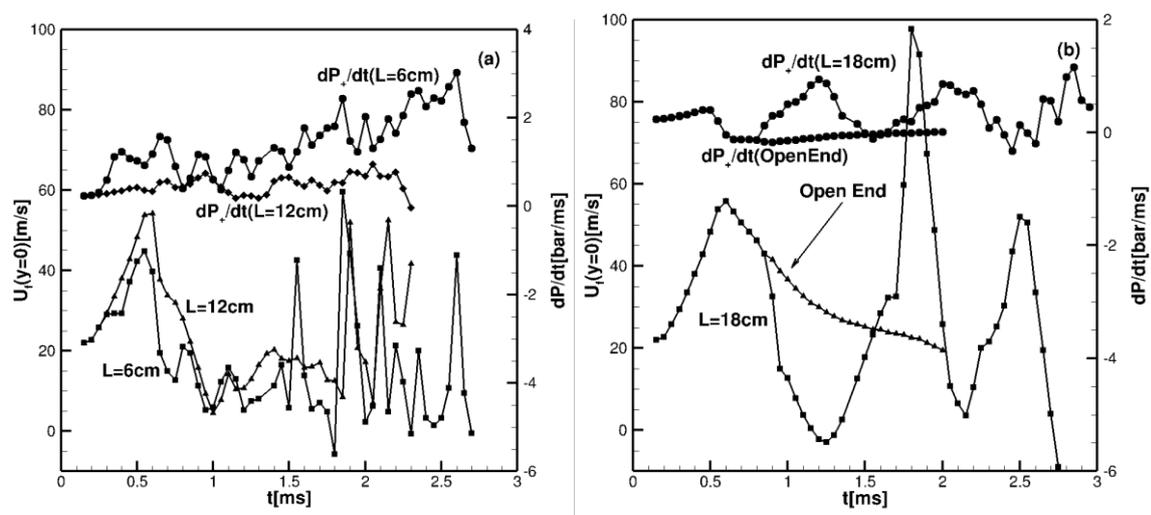

**Fig. 2: a)** The local velocity of the flame front at the tube axis for tubes $L/D = 6, 12$ and the pressure growth rates $dP_+ / dt$ at the center line just ahead of the flame as a function of time; **b)** the same for the tube with aspect ratio $L/D=18$ and for the half-open tube.

Fig. 2(a) shows that the first collision of the flame with a pressure wave in the tube with $\alpha=6$ occurs at 0.3ms, resulting in a decrease in flame velocity. After the flame skirt touches the side walls, the dynamics of the flame are different depending on the aspect ratio of the tubes. The intensity of the rarefaction wave generated by the decelerating flame can be characterized by the absolute value of flame acceleration. The effect of the first rarefaction wave, which occurs after 0.55ms when the flame surface area and flame velocity begin to decrease, can be further enhanced by the flame collision with the reflected pressure wave. Figures 1(a) and 2(a) show that up to 1ms, the flame deceleration is stronger for the tube with smaller aspect ratio, $\alpha=6$, compared to the tube with larger aspect ratio, $\alpha=12$. This trend is better seen in Figures 1(b) and 2(b) for the tube with aspect ratio $\alpha=18$ compared to a semi-open tube. From 0.55ms to 0.85ms the flame deceleration rate is the same for the tube with aspect ratio $\alpha=18$ and for the semi-open tube. But after 0.9ms, when the flame in the closed tube collides with the reflected pressure wave, the flame deceleration rate in the closed tube is much greater than in the semi-open tube.

From 0.55ms to 0.65ms, the flame decelerations are about the same in both tubes with aspect ratio $\alpha=6$ and $\alpha=12$. In the tube with aspect ratio $\alpha=6$, deceleration rate decreases at 0.65ms, but then increases at 0.8ms due to the flame collision with the reflected pressure wave. Different competing processes can affect the shape of the flame front. The decelerating flame acts as a semi-transparent piston, that in the reference frame of the flow ahead of the flame begins to move with acceleration out of the tube. It is known [34] that such a piston generates a simple rarefaction wave. The superposition of the unburned gas flow created by the rarefaction wave and the unburned gas flow created earlier by the flame in the acceleration stages leads first to the formation of a flat flame front and then to the inversion of the flame front and the formation of a tulip flame [29, 31]. A decelerating flame front can be distorted by Rayleigh-Taylor instability if small disturbances can grow significantly during the deceleration time $\Delta t$. This occurs if the increment of the RT instability, $\sigma_{RT}$, is large enough to satisfy the condition $\sigma_{RT}\Delta t >> 1$.

*3.2 Two-dimensional tube with aspect ratio $\alpha = L/D = 6$*

In this and subsequent Sections 3.3-3.5, we describe the flame dynamics in terms of the flame front velocity $U_f(y=0)$ along the centerline and the flame front velocity $U_f(y=0.4cm)$ near the tube sidewalls. This description helps to understand the evolution of the flame shape and the effects of flame collisions with pressure waves.

Figures 3(a, b) show the time evolution of: a) unburned gas velocities calculated at 0.5 mm ahead of the flame front at $y=0$ and near the tube wall, $y=0.4$cm; b) velocities of the flame front at $y=0$, $y=0.4$cm, which are related with the variation of the flame surface area $F_f$, for the tube with aspect ratio $\alpha = 6$.

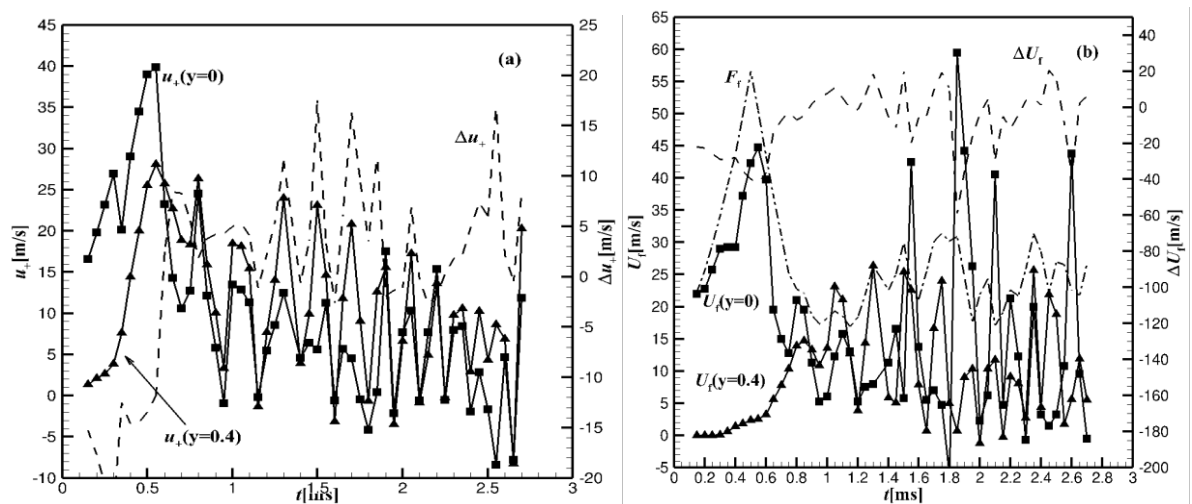

**Fig. 3(a, b):** (**a**) Temporal evolution of the unreacted flow velocities $u_+(X_f + 0.5mm)$ at 0.5 mm ahead of the flame front at the tube axis $y=0$ and near the tube wall $y=0.4cm$, and $\Delta u_+ = u_+(y=0.4cm) - u_+(y=0)$. (**b**) Temporal evolution of the flame surface area $F_f$, the flame front velocities at the tube axis $y=0$ and at $y=0.4$cm, and the difference $\Delta U_f = U_f(y=0.4cm) - U_f(y=0)$. Both figures are for aspect ratio $\alpha = 6$.

It can be seen in Fig. 3(a) that at $\approx 0.57$ms the unburned gas velocity $u_+(y=0.4cm)$ close to the tube wall exceeds the unburned gas velocity $u_+(y=0)$ at the tube axis, and their difference, $\Delta u_+$, becomes positive (dashed line). The flame skirt touches the side wall of the tube at 0.55ms, so the time interval $\Delta t=0.02$ms can be seen as the time it takes for the rarefaction wave to create the reverse flow in the unburned gas, in agreement with the estimate, $t_{RW} \approx D/a_s = 0.02$ms, used earlier. After 0.57ms the unburned gas velocity immediately ahead of the flame front is minimum at the tube axis, $y=0$, increases gradually towards the tube wall, reaches a maximum at $y=0.4cm$, decreases in the boundary layer and vanishes at the tube wall. Thus, after 0.57ms, the unburned gas velocity profile takes on a tulip shape immediately ahead of the flame front (red lines in Fig.4). In the thin flame model, each part of the flame front moves relative to the unburned mixture with a laminar flame velocity $U_{fL}$ and is simultaneously entrained by the flow of unburned gas at a local flow velocity $u_+(x,y)$ immediately ahead of that part of the flame front, $U_f(x,y) = U_{fL} + u_+(x,y)$. Therefore, the flame shape "replicates" the shape of the unburned flow velocity profile and also takes on a tulip shape. This can be seen in Fig. 3(b), which shows the time evolution of the flame front velocities at the tube axis, $y=0$, near that tube wall at $y=0.4cm$, their difference $\Delta U_f$, and the flame surface area $F_f$. It can be seen that the flame velocities are closely related to the variations in the flame surface area. A small phase difference appears at later stages during the flame collisions with the pressure waves.

After 1.2ms the pressure increases from the initial value of 1atm to over 2.5atm, and the flame front becomes more than twice as thin. Fig. 3(b) shows that every collision of the flame with the pressure wave reflected from the right closed end of the tube leads to an additional short, $\Delta t \approx 0.2\,ms$ phase of the flame deceleration. This can lead partly to the formation a deeper tulip flame (tulip with longer petals), and at the same time, to a larger (negative) flame acceleration, which favors the development of short wavelength, $\lambda \sim 0.2D$, perturbations of the RT instability, for which $\sigma_{RT}\Delta t \gg 1$. This trend continues until bulges on the petals of the flame tulip, caused by the RT instability, move toward the tube axis, completely destroying the tulip-shaped flame and the flame front becomes "wavy flat". The oscillations of the flame surface area which are seen in Fig. 3(b) are the result of complex competing processes. The development of a tulip-shaped flame front increases the surface area of the flame, but can be slowed down by the collision of the flame with pressure waves or enhanced by the growing bulges on the tulip petals caused by RT instability.

Figure 4 shows computed schlieren images and streamlines at selected times during tulip flame formation in the tube with an aspect ratio $\alpha = 6$. Red lines show the axial velocity profiles in the unburned gas at 0.5mm ahead of the flame; dashed red lines show the location of the corresponding velocity profiles. While the velocity profile in the unburned gas ahead of the flame is initially convex (image at 0.25 ms), it takes on a concave tulip shape just ahead of the flame front as the flame decelerates. Far ahead from the flame, the velocity profile in the unburned flow remains convex: the flow is nearly uniform in the bulk with velocity decreasing within a thin boundary layer. The oblique pressure waves seen in Fig. 4 were generated earlier by the convex flame during the acceleration phases and propagate through the tube, repeatedly colliding and reflecting off the side walls, reflecting off the right end of the tube, and moving back and forth along the tube. Figure 2a shows that in a tube with aspect ratio $\alpha = 6$ several collisions of the flame with the pressure waves at ≈0.6ms and ≈0.8ms intensify the rarefaction wave, resulting in the formation of a tulip flame at ≈1.0ms. It can be seen in Fig. 4 that the flame front inverts from the initially convex shape to a concave tulip shape in a short time of ≈0.2ms. The increment of the RT instability is

$$\sigma_{RT} = \sqrt{Akg} , \qquad (20)$$

where $A = (\rho_u - \rho_b)/(\rho_u + \rho_b)$ is Attwood number, $k = 2\pi/\lambda$ is a wave number. Calculating $\sigma_{RT}$ for the flame accelerations $g = |a|$, before 1.2 ms (see Table 1) and perturbations with wavelength $\lambda \sim D$, which conventionally could lead to the flame front inversion, we obtain that for $\Delta t \approx 0.2 \text{ms}$

$$\sigma_{RT} \Delta t < 1. \tag{21}$$

This means that the RT instability is not involved in tulip flame formation up to 1.2 ms. The same is true for the DL instability. Using Eq. (1), we find that the characteristic time for the development of the DL instability for perturbations with wavelength $\lambda \sim D$ is 0.6ms, which is noticeably longer than the time of flame front inversion.

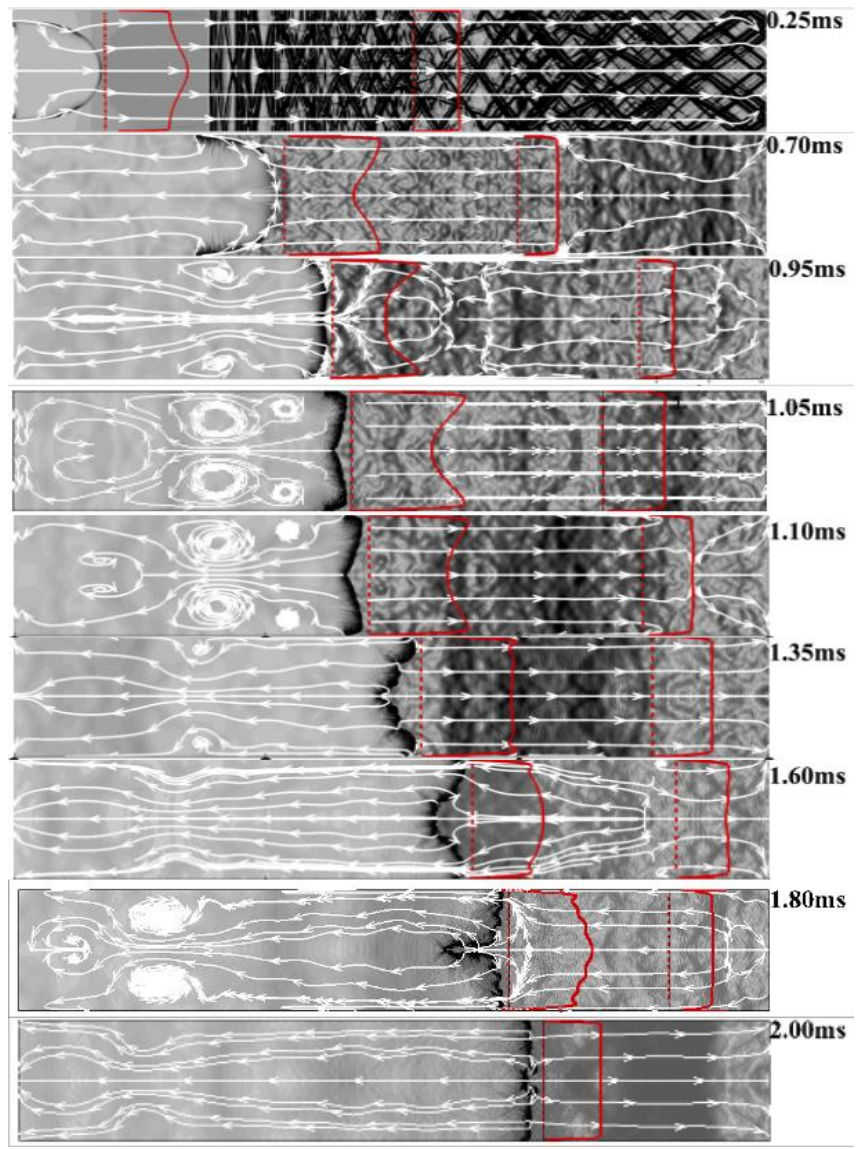

**Fig. 4**. Sequences of computed schlieren images and stream lines for a flame in the tube with aspect ratio $\alpha = 6$. Velocity profiles in the unburned gas (red lines) are shown at 0.5mm ahead of the flame and far ahead of the flame.

Figures 2(a) and 3(b) show that after ≈1.35 ms the flame collisions with reflected pressure waves lead to a significantly higher acceleration of the flame, so that $\sigma_{RT}\Delta t \gg 1$ for short wavelength perturbations $\lambda \sim 0.2D$, which is sufficient for the development of the Rayleigh-Taylor instability leading to the distorted tulip flame, which is seen in Fig. 4 at 1.35 ms, 1.6 ms and 1.8 ms.

### *3.3 Two-dimensional tube with aspect ratio $\alpha = L/D = 12$*

Figure 2a shows that since the flame in the tube with an aspect ratio $\alpha = 12$ does not collide with the reflected pressure wave in the acceleration phase, it reaches a higher velocity before the flame skirt touches the tube wall than the flame in the tube with smaller aspect ratio $\alpha = 6$ discussed in Sec. 3.2. Figure 5 shows computed schlieren images and streamlines at selected times during tulip flame formation in the tube with aspect ratio $L/D = 12$.

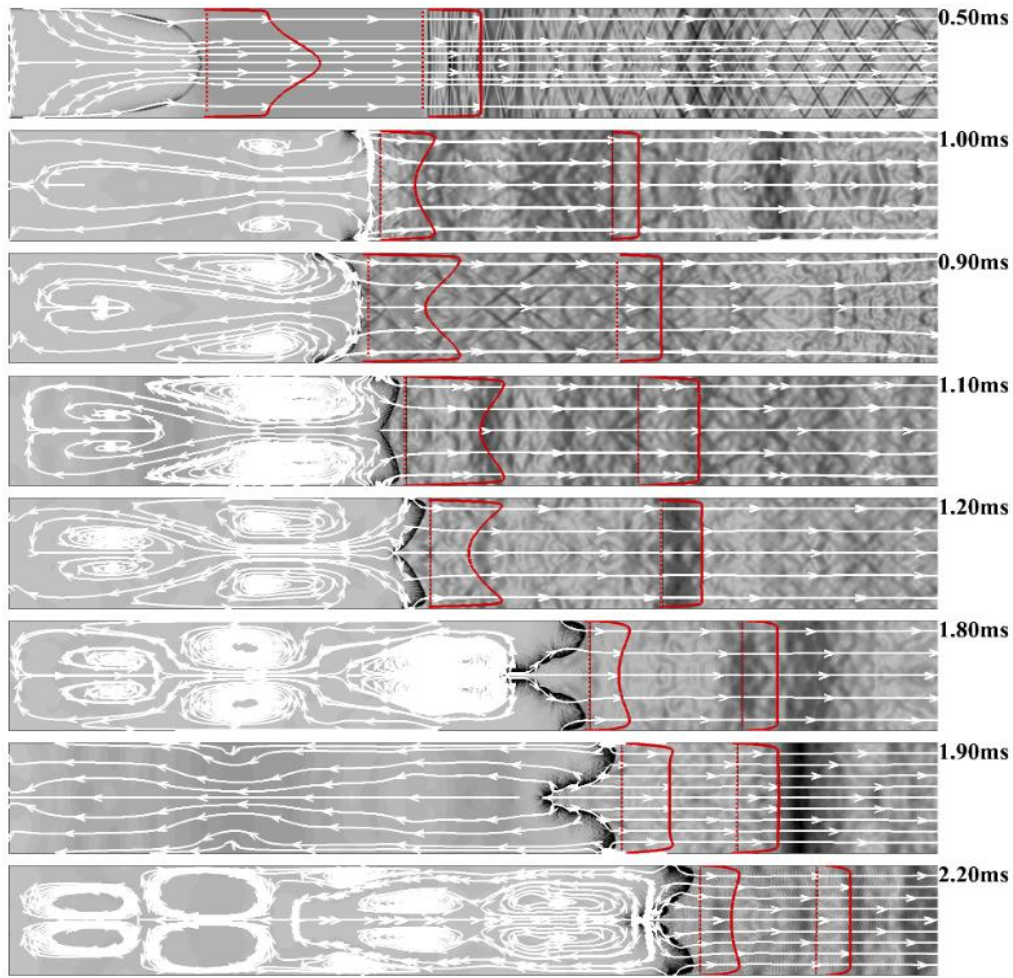

**Fig. 5.** Sequences of computed schlieren images and stream lines for a flame in the tube with aspect ratio $L/D=12$. Velocity profiles in the unburned gas (red lines) are shown at 0.5mm ahead of the flame and far ahead of the flame.

It can be seen that the classic tulip shaped flame is forming later (at $\approx 1.2$ms) than for the flame in the tube with aspect ratio $\alpha=6$.

Figures 6(a, b) show the time evolution of: **a)** unburned gas velocities calculated at 0.5 mm ahead of the flame front at $y=0$ and near the wall, at $y=0.2$cm and $y=0.4$cm, and the difference $\Delta u_+ = u_+(y=0.4cm) - u_+(y=0)$; **b)** velocities of the flame front at $y=0$ and at $y=0.4$cm, the difference $\Delta U_f = U_f(y=0.4cm) - U_f(y=0)$ and the flame surface area $F_f$.

Figure 6(b) shows the time evolution of the flame front velocities at the tube axis, $y=0$, near the tube wall, $y=0.4cm$, and the difference $\Delta U_f = U_f(y=0.4cm) - U_f(y=0)$. The tulip-shaped velocity profile in the unburned gas is formed later than in the tube with an aspect ratio $\alpha=6$: $\Delta u_+ > 0$ at $\approx 0.75$ ms. Therefore, the inversion of the flame front also starts later, at $\approx 0.93$ms, when $\Delta U_f = U_f(y=0.4cm) - U_f(y=0)$ becomes positive.

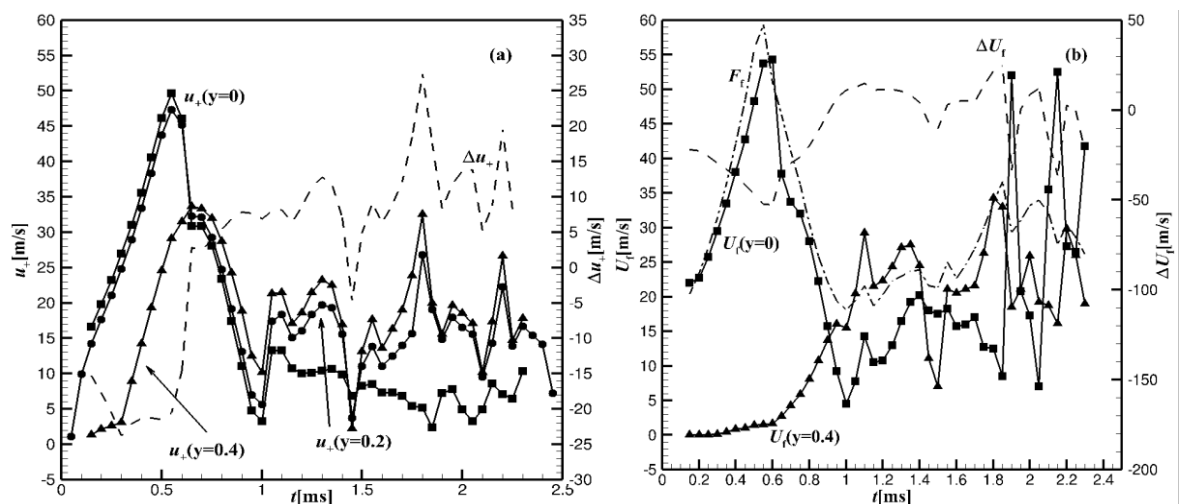

**Fig. 6(a, b):** (a) Time evolution of the unreacted flow velocities $u_+(X_f + 0.5mm)$ at 0.5 mm ahead of the flame front at $y=0$ and $y=0.4cm$, and $\Delta u_+ = u_+(y=0.4cm) - u_+(y=0)$. (b) Time evolution of the flame surface area $F_f$ and the flame front velocities at $y=0$ and $y=0.4cm$, calculated overpressure $\Delta P$, and $\Delta U_f = U_f(y=0.4cm) - U_f(y=0)$; aspect ratio $L/D=12$.

Figure 6(b) shows that the collisions of the flame with pressure waves at 1.82ms and at 2.05ms result in a large enough acceleration (see Table 1) for the RT instability, leading to a distorted tulip flame. In this case the pressure does not rise as high as in the tube with smaller aspect ratio, $\alpha = 6$, and therefore the distorted tulip flame is not as pronounced as in the shorter tube. After 1.2 ms, the tulip shape formation continues briefly, the tulip becomes deeper, the tulip petals continue to grow, concurrent with the formation of the distorted tulip flame, and after 1.35 ms, the subsequent flame shape changes are determined only by the RT instabilities.

### 3.4 Two-dimensional tube with aspect ratio $\alpha = L/D = 18$

Figure 7(a, b) show the temporal evolution of: **a)** the unburned gas velocity $u_+(y=0)$ at the tube axis and near the tube wall, $u_+(y=0.4cm)$, calculated at 0.5 mm ahead of the flame front for the tube with $\alpha = 18$ and the difference $\Delta u_+ = u_+(y=0.4cm) - u_+(y=0)$; **b)** the time evolution of the flame front velocities $U_f(y=0)$ and $U_f(y=0.4cm)$, their difference $\Delta U_f = U_f(y=0.4cm) - U_f(y=0)$, and the flame surface area $F_f$.

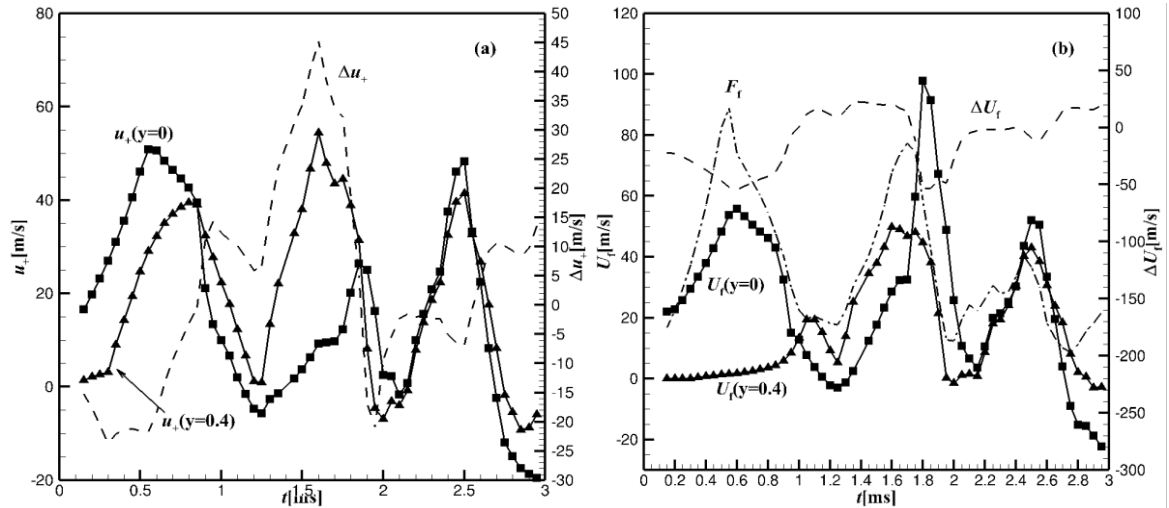

**Fig. 7(a, b)**: (a) Temporal evolution of the unreacted flow velocities $u_+(X_f + 0.5mm)$ at 0.5 mm ahead of the flame front at $y=0$ and $y=0.4cm$, and $\Delta u_+ = u_+(y=0.4cm) - u_+(y=0)$. (b) Temporal evolution of the flame surface area $F_f$ and the flame front velocities at $y=0$ and $y=0.4cm$, and $\Delta U_f = U_f(y=0.4cm) - U_f(y=0)$; aspect ratio $\alpha = 18$.

Figure 8 shows computed schlieren images and streamlines at selected times during tulip flame formation and later in the 2D tube with aspect ratio $\alpha = 18$.

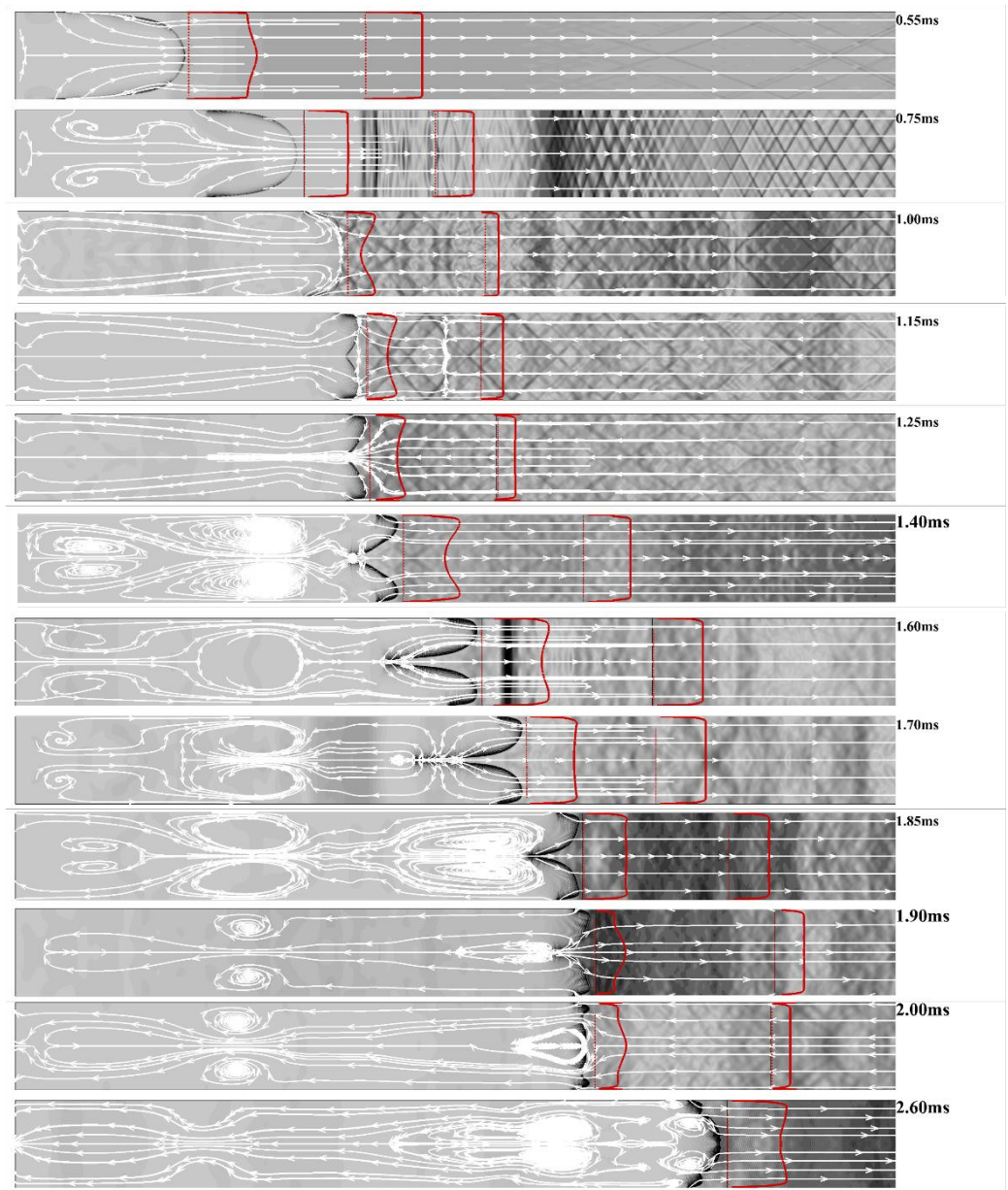

**Fig. 8**. Sequences of computed schlieren images and streamlines during tulip flame formation in the channel $L/D=18$. The axial velocity profiles in the unburned gas (red lines) are shown at 0.5 mm ahead of the flame front and far ahead from the flame.

It can be seen in Fig. 8 that the tulip-shaped flame is formed due to the rarefaction wave caused by the decelerating flame as the flame surface area begins to decrease after the flame skirt touched the tube sidewalls. At ≈0.85ms the rarefaction was enhanced by the flame collision with reflected pressure wave (Fig. 2b). In Fig. 7b and 7a it is seen that the flame surface area continues to decrease, but the velocities $U_f(y=0)$ and $u_+(y=0)$ decrease more rapidly. After $\approx 1.0$ ms, the flame surfave area increases during the tulip shape development, and therefore the flame velocity also increases up to $\approx 1.70$ms. At $\approx 1.75$ms the collision of the flame with relatively strong reflected pressure wave slowdowns the flame leading to the development of the RT instability and to the appearance of a distorted tulip flame. In Fig. 7.b it can be seen that in the time interval from t$\approx 0.95$m up to 1.75ms $\Delta U_f$ is positive and the classical tulip flame is formed up to $\approx 1.70$ms. The subsequent scenario is similar to that for a flame in a tube with aspect ratio $\alpha = 12$. After the flame collides with a reflected pressure wave at $\approx 1.8$ms, a distorted tulip flame is formed due to the RT instability and the flame surface area increases. Later, the bulges begin to move toward the axis, $\Delta U_f \leq 0$ between 1.8 ms and 2.4ms, and the flame speed decreases.

*3.5 Two-dimensional semi-open tube*

Figure 9(a, b) shows the time evolution of: **a)** the unburned gas velocity at the tube axis, and near the tube wall, at $y = 0.4 cm$, calculated 0.5 mm ahead of the flame front; **b)** the time evolution of the flame front velocities at the same points and the flame surface area. Because there are no collisions of the flame with reflected pressure waves that could intensify tulip flame formation, flame front inversion takes longer time and the tulip flame shape is not as pronounced as in tubes with both ends closed. The flame slows down with acceleration (Table 1) $a = -4.7 \cdot 10^6$ cm/s from 0.55ms until 1.1ms and with acceleration $a = -1.1 \cdot 10^6$ cm/s from 1.1ms until $\approx 1.7$ms. During both time intervals 0.55ms and 0.9ms, $\sigma_{RT}\Delta t \leq 1$, that is insufficient for the development of Rayleigh-Taylor instability with the wavelength $\lambda \sim D$. Therefore, the RT instability does not participate in the formation of the tulip flame. The "distorted" tulip flame, which can be seen in the schlieren images at 1.8 and at 2.0ms in Fig. 10, is due to the Darrieus-Landau instability. In this case the increment of the DL instability

for the development of short-wavelength perturbations $\lambda \approx 0.2 cm$, seen at the image at 2.0ms in Fig. 10, is sufficiently large, $\sigma_{DL} \approx 20$, to satisfy the condition $\sigma_{DL}\Delta t \gg 1$ for $\Delta t = 0.2\text{ms}$.

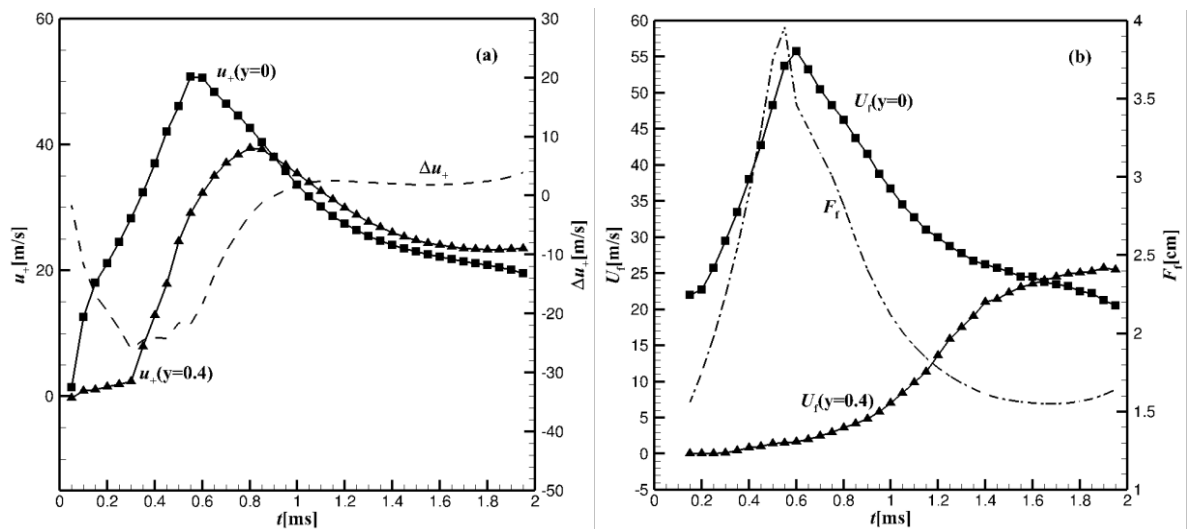

**Fig. 9(a, b)**: (a) Evolution of the unburned gas velocity at the tube axis, $y = 0$, and near the tube wall, $y = 0.4 cm$, calculated at 0.5 mm ahead of the flame front; (b) time evolution of the flame front velocities at $y = 0$ and $y = 0.4 cm$; semi-open tube.

Figure 10 shows computed schlieren images, streamlines, and the velocity profiles in the unburned gas during tulip flame formation for a flame ignited at the left closed end of the $D = 1 cm$ wide tube propagating to the opposite open end.

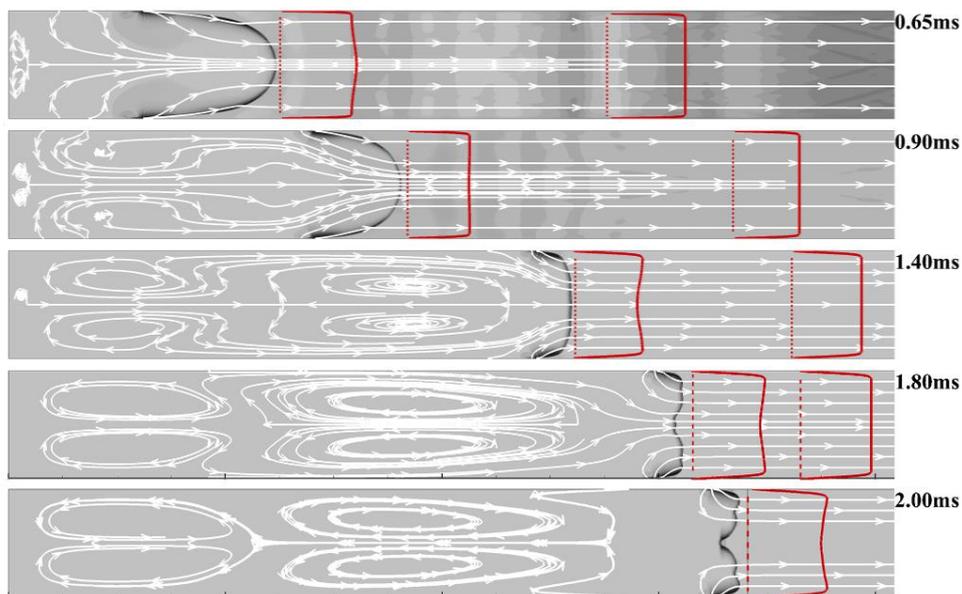

**Fig. 10.** Sequences of computed schlieren images, streamlines during tulip flame formation in a half-open tube. The velocity profiles in the unburned gas are shown at 0.5mm in front of the flame front, with their locations indicated by the dashed lines, and far ahead of the flame.

Table 2

Flame accelerations in tubes with aspect ratios α= 6, 12, 18, and in a semi-open tube.

| α=L/D=6 | | α=L/D=12 | | α=L/D=18 | | Semi-open tube | |
|---|---|---|---|---|---|---|---|
| t/ms | $g = \|a\|$ | t/ms | $g = \|a\|$ | t/ms | $g = \|a\|$ | t/ms | $g = \|a\|$ |
| 0.55-0.6 | $1.3\times10^5$m/s$^2$ | 0.6-0.65 | $2.0\times10^5$m/s$^2$ | 0.6-0.85 | $4.7\times10^4$m/s$^2$ | 0.6-1.25 | $4.7\times10^4$m/s$^2$ |
| 0.6-0.65 | $2.5\times10^5$m/s$^2$ | 0.75-1.0 | $1.3\times10^5$m/s$^2$ | 0.85-0.97 | $2.6\times10^5$m/s$^2$ | 1.2-1.8 | $1.1\times10^4$m/s$^2$ |
| 0.8-0.95 | $1.4\times10^5$m/s$^2$ | 1.10-1.20 | $3.3\times10^4$m/s$^2$ | 0.97-1.25 | $7.2\times10^4$m/s$^2$ | | |
| 1.1-1.2 | $6.3\times10^4$m/s$^2$ | 1.9-1.55 | $3.3\times10^5$m/s$^2$ | 1.8-2.05 | $4.2\times10^5$m/s$^2$ | | |
| 1.55-1.65 | $3.7\times10^5$m/s$^2$ | 1.95-2.0 | $3.5\times10^5$m/s$^2$ | 2.5-2.75 | $3.1\times10^5$m/s$^2$ | | |
| 1.85-2.0 | $4.2\times10^5$m/s$^2$ | 2.15-2.2 | $2.5\times10^5$m/s$^2$ | | | | |

## *3.6 Three-dimensional channel*

Flame dynamics and tulip flame formation in a 3D rectangular channel are qualitatively similar to those in a 2D channel. The most significant difference is the much higher velocity and the rates of flame acceleration and deceleration due to the increase or decrease in flame surface area [29, 30], so that the maximum flame velocity achieved in the acceleration phase is almost twice as high in the 3D case than in the 2D case. Figure 11(a, b) shows: **a)** the time evolution of the unburned flow velocities at 0.5mm ahead of the flame at the cross section ($x, y, z = 0$), at the axis $y = z = 0$, near the tube wall, $y = 0.3$cm, and $\Delta u_+ = u_+(y = 0.3\text{cm}) - u_+(y = 0)$, for the flame propagating in the rectangular channel of length $L = 6$cm and cross section $D \times D = 1\text{cm}^2$; **b)** the time evolution of the flame surface area $F_f$, the flame front velocities at the tube axis $U_f(y = z = 0)$ and near the walls $U_f(y = 0.3\text{cm})$, and pressure $P$ just ahead of the flame.

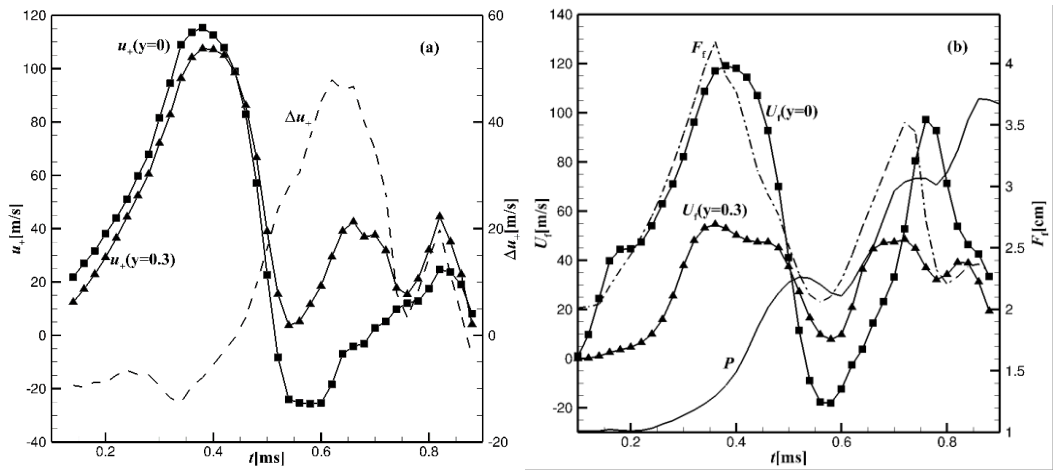

**Fig. 11(a, b):** (a) The time evolution of the unburned flow velocities 0.5mm ahead of the flame front at the tube axis $y = z = 0$ and near the wall, $y = 0.3\,cm$; (b) the flame surface area, $F_f$, local velocities of the flame front at the tube axis $y = z = 0$ and at near the wall $y = 0.3\,cm$. Flame propagates in the tube with both ends closed, $L/D = 6$, $D \times D = 1\,cm^2$.

Figure 12 shows a sequence of numerical schlieren images, stream lines and velocity profiles in the unburned gas in the cross section ($x, y, z = 0$), which locations ahead of the flame front are shown by the dashed lines.

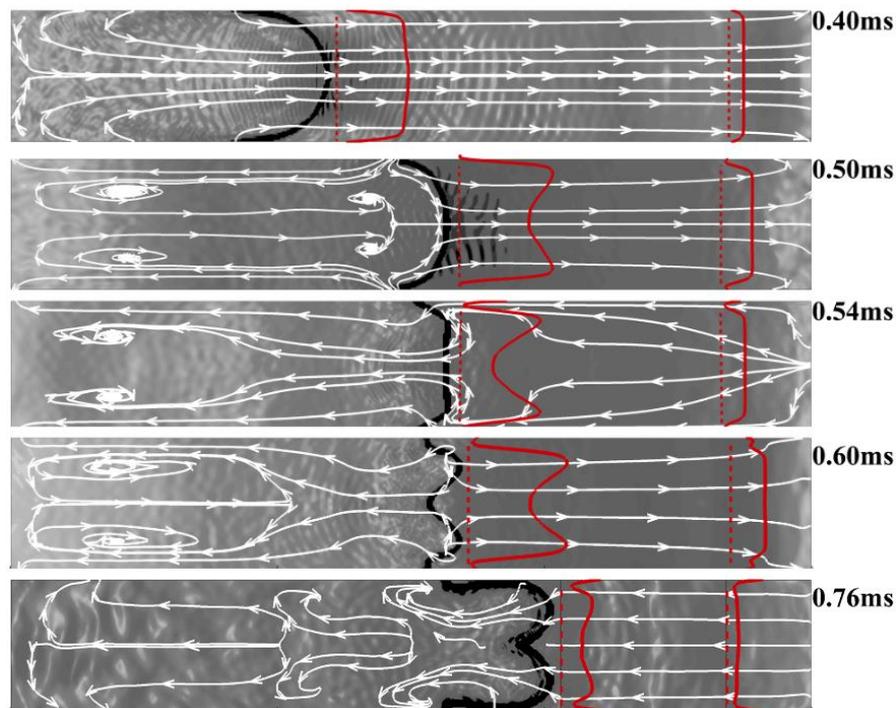

**Fig. 12.** Sequence of schlieren images, stream lines and velocity profiles in the unburned gas for the flame propagating in 3D tube with both ends closed, $L/D = 6$, cross section $D \times D = 1\,cm^2$, shown is the cross section ($x, y, z = 0$).

As mentioned above, the acceleration and the rarefaction wave intensity are much higher in the case of the 3D flame, and the tulip flame is formed faster than in case of 2D flames.

### 4. Discussions and conclusions

The earlier stages of hydrogen-air flame propagation in closed and semi-open two- and three-dimensional tubes and the mechanism of tulip flame formation have been studied, with special emphasis on the effects of pressure waves on the formation of tulip and distorted tulip flames. Numerical simulations of 2D stoichiometric hydrogen-air premixed flames in closed and semi-open tubes were performed by solving the 2D fully compressible Navier-Stokes equations coupled with detailed chemical kinetics using the WENO fifth-order finite difference code [36], and the LES thickened flame method [43] was used for 3D flame simulations.

The major conclusion of this paper is that the physical process that causes the inversion of the flame front and the formation of tulip flames are the rarefaction waves generated by the flame during the deceleration phases. The first rarefaction wave occurs when the flame skirt touches the side walls of the tube, the flame surface area begins to decrease and with it the flame velocity. In the reference frame of the unburned gas flow, the flame front can be viewed as a semi-transparent piston that begins to move out of the tube with acceleration, creating a simple rarefaction wave that generates a reverse flow of unburned gas. The superposition of the forward flow of unburned gas previously created by the accelerating flame and the reverse flow created by the rarefaction wave during flame deceleration leads to the formation of a tulip-shaped axial velocity profile of the unburned gas in the immediate vicinity of the flame front. In the theoretical model of an infinitely thin flame, the velocity of any point on the flame front is the sum of the laminar velocity of the flame relative to the unburned gas $\vec{U}_{fL}$ and the velocity of the unburned gas $\vec{u}_+(\vec{r})$ immediately ahead of the flame front at which that portion of the flame front is entrained by the unburned gas. This means that if the axial velocity profile of the unburned gas becomes tulip-shaped, the flame front will also become tulip-shaped. Although the condition $\vec{U}_f = \vec{U}_{fL} + \vec{u}_+$ is formally valid for the theoretical model of an infinitely thin flame, simulations performed for a flame with real thickness of the flame front have shown that it is satisfied with very good accuracy [31]. This scenario is universal and works as in the case where the flame deceleration is due to the reduction of the flame surface area and in the case where it is the result of the flame collision with the pressure waves.

It should be noted that the flow of unburned gas produced by a rarefaction wave generated by a decelerating flame is different from the gas flow produced by the classical rarefaction wave when a flat piston begins to move out of the tube with acceleration [34]. In the classical problem, the flow produced by a flat piston is one-dimensional. The solution for the flow produced by the rarefaction wave can be obtained using the Riemann solution for a simple one-dimensional traveling wave. It follows that the velocity in the flow between the piston and the rarefaction wave front is everywhere directed toward the piston acceleration, decreases monotonically, and disappears toward the rarefaction wave front. In our case the problem is multidimensional (2D or 3D) and does not have a simple analytical solution. Nevertheless, it is obvious that near the piston surface, the magnitude of the axial velocities in the reverse flow produced by a decelerating convex shaped piston (a decelerating finger-shaped flame front) is maximum at the tube axis and decreases toward the tube walls. Therefore, in the laboratory reference frame, the unburned gas flow, which is a superposition of the flow established in the flame acceleration phase and the flow created by the rarefaction wave, acquires a tulip-shaped velocity profile near the flame front.

For a given tube width, the tulip flame will form faster in the tube with a smaller aspect ratio, because in a shorter tube there are more flame-pressure wave collisions, which amplify the process of tulip flame formation. The semi-open tube can be considered as a limiting case of a tube with infinite aspect ratio, where the tulip flame formation takes the longest time. At the same time, the flame speed in a tube with a higher aspect ratio is higher at the moment when the flame skirt touches the tube walls (Fig. 1a). This can result in a stronger flame deceleration during the flame collision with reflected pressure waves after the formation of the classical tulip flame, which can be sufficient for the development of the Rayleigh-Taylor instability leading to a distorted tulip flame.

The formation of a deeper tulip-shaped flame with bumps on the tulip petals results in an increase in flame surface area and an increase in the flame velocity. Subsequent flame collisions with reflected pressure waves can further increase the size of the bumps, creating a flat corrugated flame surface and then an accelerating flame with a convex corrugated flame front (Fig. 8).

Another conclusion is that the inversion of the flame front and the tulip flame formation occur much faster for the 3D flame than for the 2D flame in a tube of the same width and aspect ratio. Although the 2D modeling results are qualitatively similar to those observed in experiments, they should be interpreted with caution because there are significant quantitative differences between 2D flames and real, experimental 3D flames. In particular, the characteristic transition times from spherical to finger-shaped flames and the time when the flame skirt touches the sidewall, obtained in [45] using a simple two-dimensional analytical model in an attempt to "improve" and quantify the predictions of the geometric model developed by Clanet and Searby [17], are not applicable to three-dimensional experimental flames. It is therefore not surprising that large discrepancies were found between the characteristic times measured by Shen et al. [46] and those predicted by the 2D model [45].

It is known that despite the efficient technique such as adaptive mesh refinement (AMR) providing adequate resolution of flame and boundary layers, chemical kinetics remains a computational bottleneck in reactive CFD simulations. Comparison of 2D DNS results for tulip flame formation in a closed 2D tube with $L = 6\,\text{cm}$ and aspect ratio $\alpha = 6$, with 2D LES simulations using thickened flame model with resolution $dx = 125\,\mu m$ shows that the thickened flame model gives reasonably good values of flame velocities during tulip flame formation, up to 0.7ms. However, further flame evolution related to flame pressure wave collisions is not well modeled by the thickened flame model. Also, our modeling of flame velocity using a thickened flame model for experimental conditions [44, 47] has shown satisfactory agreement with experimentally measured flame velocity during tulip flame formation, but not during later stages of flame propagation. Therefore, adaptive mesh refinement (AMR) and dynamic load balancing (DLB) algorithms based on the formation of groups of processors that share the load of the detailed chemical model [48, 49] should be used to reliably simulate the tulip flame under experimental conditions.

**Declaration of competing interest**

The authors declare that they have no known competing financial interests or personal relationships that could have appeared to influence the work reported in this paper.

**Acknowledgements**

The authors thank the support of this work provided by Olle Engkvists Foundation (grant No.232-0212, project no 31005258) and National Natural Science Foundation of China(CQ).

**References**

[1] O.C. Ellis, R.V. Wheeler, The movement of flame in closed vessels. J. Chem. Soc. 127 (1925) 764–772.

[2] O. Ellis, Flame movement in gaseous explosive mixtures. Fuel Sci.7 (1928) 502–508.

[3] G.D. Salamandra, T.V. Bazhenova, I.M. Naboko, Formation of a detonation wave in gas combustion in tubes. Symp. (Int.) Combust., 7 (1959) 851–855.

[4] G. H. Markstein, A shock-tube study of flame front-pressure wave interaction. Symp. (Int.) Combust., Proc. Combust. Inst. 6 (1957) pp. 387–398.

[5] G. H. Markstein, Experimental studies of flame-front instability. In: Non-steady Flame Propagation. G.H. Markstein, ed. Pergamon Press, New York, 1964.

[6] J.C. Leyer, N. Manson, Proc. Development of vibratory flame propagation in short closed tubes and vessels. Proc. Combust. Inst. 13 (1971) pp. 551–558.

[7] R. Starke, P. Roth, An experimental investigation of flame behavior during cylindrical vessel explosions. Combust. Flame, 66 (1986) pp. 249-259.

[8] D. Dunn-Rankin and R.F. Sawyer, Tulip flames: Changes in shapes of premixed flames propagating in closed tubes, Experiments in Fluids 24 (1998), pp. 130–140.

[9] M. Gonzalez, R. Borghi, A. Saouab, Interaction of a flame front with its self-generated flow in an enclosure: The "tulip flame" phenomenon. Combust. Flame, 88 (1992) pp. 201–220.

[10] B. N'Konga, G. Fernandez, H. Guillard, B. Larrouturou, Numerical investigations of the tulip flame instability - comparisons with experimental results. Combust. Sci. Tech., 87 (1993) pp. 69–89.

[11] M. Matalon, J.L. McGreevy, The initial development of the tulip flame. Proc. Combust. Inst. 25 (1994) pp. 1407–1413.

[12] J.L. McGreevy, M. Matalon, Hydrodynamic instability of a premixed flame under confinement. Combust. Sci. Technol., 100 (1994) pp. 75–94.

[13] J.W. Dold, G. Joulin, An evolution equation modeling inversion of tulip flames. Combust. Flame, 100 (1995) pp. 450-456.

[14] M. Gonzalez, Acoustic Instability of a Premixed Flame Propagating in a Tube. Combust. Flame, 107 (1996) pp. 245-259.

[15] V.V. Bychkov, M. A. Liberman, On the Stability of a Flame in a Closed Chamber. Phys. Rev. Letters, 78 (1997) pp. 1371 -1734.

[16] M.A. Liberman, V.V. Bychkov, S.M. Golberg, L.E Eriksson, Numerical Study of Curved Flames under Confinement. Combust. Sci. Technol. 136 (1998) pp. 221-251.

[17] C. Clanet, G. Searby, On the 'tulip flame' phenomenon, Combust. Flame, 105 (1996) 225–238.

[18] D. Dunn-Rankin, The interaction between a laminar flame and its self-generated flow. Ph.D. Dissertation, University of California, Berkeley, 1985.

[19] M. Matalon, P. Metzener, The propagation of premixed flames in closed tubes. J. Fluid Mech., 336 (1997) pp. 331–350.

[20] T. Kratzel, E. Pantow, M. Fischer, On the transition from a highly turbulent curved flame into a tulip flame. Int. J. Hydrogen Energy, 23 (1998) pp. 45-51

[21] A.K. Kaltayev, U.R. Riedel, J. Warnatz, The Hydrodynamic Structure of a Methane-Air Tulip Flame. Combust. Sci. Tech. 158 (2000) pp. 53-69.

[22] S. Kerampran, D. Desbordes and B. Veyssière, Study of the mechanisms of flame acceleration in a tube of constant cross section. Combust. Sci. Technol. 158 (2000) pp. 71–91.

[23] P. Metzener, M. Matalon, Premixed flames in closed cylindrical tubes. Combust. Theor. Model., 5 (2001) pp. 463–483.

[24] H. Xiao, D. Makarov, J. Sun, V. Molkov, Experimental and numerical investigation of premixed flame propagation with distorted tulip shape in a closed duct. Combust. Flame, 159 (2012) pp. 1523–1538.

[25] R. Kiran, I. Wichman, N. Mueller, On combustion in a closed rectangular channel with initial vorticity. Combust. Theory Modelling, 18 (2014) pp. 272-294.

[26] H. Xiao, R.W. Houim, E.S. Oran, Formation and evolution of distorted tulip flames. Combust. Flame, 162 (2015) pp. 4084–4101.

[27] D. Dunn-Rankin, Combustion Phenomena: Selected Mechanisms of Flame Formation, Propagation and Extinction. Ed. J. Jarosinski and B. Veyssiere. CRC Press, Boca Raton, FL. 2009.

[28] G. Searby, Combustion Phenomena, CRC Press, Boca Raton, FL, 2009.

[29] M. A. Liberman, C. Qian, C. Wang, Dynamics of flames in tubes with no-slip walls and the mechanism of tulip flame formation. Combust. Sci. Technol. 195 (2023) 1637-1665.

[30] M. A. Liberman, Combustion Physics: Flames, Detonations, Explosions, Astrophysical Combustion and Inertial Confinement Fusion. Springer 2021. ISBN 978-3-030-85138-5.

[31] Chengeng Qian, Mikhail A. Liberman, On the mechanism of "tulip flame" formation: The effect of ignition sources. Phys. Fluids 35 (2023) pp. 116122 (1-13); doi: 10.1063/5.0174234

[32] B. Ponizy, A. Claverie, B. Veyssière, Tulip flame - the mechanism of flame front inversion. Combust. Flame, 161 (2014) 3051–3062.

[33] Ya.B. Zeldovich, G.I. Barenblatt, V.B. Librovich, G.M. Makhviladze, Mathematical theory of combustion and explosion. New York: Consultants Bureau; 1985.

[34] L.D. Landau, E.M. Lifshitz, Fluid Mechanics, Volume 6, second ed. Pergamon Press, Oxford, 1989.

[35] X. Shen, C. Zhang, G. Xiu, H. Zhu, Evolution of premixed stoichiometric hydrogen/air flame in a closed duct. Energy 176 (2019) pp. 265-271.

[36] G-S. Jiang and C-W. Shu, J. Efficient implementation of weighted ENO schemes. J. Comput. Physics, 126 (1996) 202–228.

[37] A. Kéromnès, W.K. Metcalfe, K.A. Heufer, N. Donohoe, A.K. Das, C.-J. Sung, J. Herzler, C. Naumann, P. Griebel, O. Mathieu, M.C. Krejci, E.L. Petersen, W.J. Pitz, and H.J. Curran, An experimental and detailed chemical kinetic modeling study of hydrogen and syngas mixture oxidation at elevated pressures, Combust. Flame 160 (2013) 995–1011.

[38] C. Wang, C. Qian, J. Liu, M. Liberman, Influence of chemical kinetics on detonation initiating by temperature gradients in methane/air. Combust. Flame, 197 (2018) 400-415.

[39] M. Liberman, C. Wang, C. Qian, J. Liu, Influence of chemical kinetics on spontaneous waves and detonation initiation in highly reactive and low reactive mixtures. Combust. Theory and Modelling, 23 (2019) 467-495.

[40] R. Byron Bird, Warren E. Stewart, Edwin N. Lightfoot, Transport phenomena. Second ed. John Wiley and Sons, Inc. 2002.

[41] C. R. Wilke, A viscosity equation for gas mixtures. J. Chem. Phys., **18** (1950) pp. 517-519.

[42] O. Colin, F. Ducros, D. Veynante, T. Poinsot, A thickened flame model for large eddy simulations of turbulent premixed combustion. Phys. Fluids, 12 (2000) 1843-1863.

[43] E. Garnier, N. Adams, P. Sagaut, Large Eddy Simulation for Compressible Flows. Springer Science & Business Media, Springer 2009.

[44] T. Poinsot, D. Veynante, Theoretical and numerical combustion. RT Edwards, Inc., 2005.

[45] V. Bychkov, V.y. Akkerman, G. Fru, A. Petchenko, L.-E. Eriksson, Flame acceleration in the early stages of burning in tubes, Combust. Flame 150 (2007) pp. 263-276.

[46] X. Shen, C. Zhang, G. Xiu, H. Zhu, Evolution of premixed stoichiometric hydrogen/air flame in a closed duct. Energy 176 (2019) pp. 265-271.

[47] X. Shen, J. Xu, H. Zhu, Phenomenological characteristics of hydrogen/air premixed flame propagation in closed rectangular channels. Renewable Energy 174 (2021) pp. 606-615.

[48] B. Tekgül, P. Peltonen, H. Kahila, O. Kaario, V. Vuorinen, Comput. Phys. Commun. 267 (2021) 108073.

[49] J. Sun, Y. Wang, B. Tian, Z. Chen, detonationFoam: An open-source solver for simulation of gaseous detonation based on OpenFOAM. Computer Physics Communications, 292 (2023) pp. 108859